\documentclass[twocolumn, twoside]{IEEEtran}
\usepackage{amsmath,amssymb,amsfonts}
\usepackage{graphicx}
\usepackage{lipsum}
\usepackage{authblk}
\usepackage[hidelinks]{hyperref}
\usepackage{algorithm}
\usepackage{graphicx,color}
\usepackage{textcomp}
\usepackage{stfloats}
\usepackage{url}
\usepackage{verbatim}
\usepackage{graphicx}
\usepackage{cite}
\usepackage{booktabs}
\usepackage{cleveref}
\usepackage{algpseudocode}
\usepackage{multirow}

\usepackage{fancyhdr}
\fancypagestyle{plain}{%
  \fancyhf{}
  \fancyfoot[C]{\thepage}
  
}

\begin{document}

\title{Spatial Impulse Response Analysis and Ensemble Learning for Efficient Precision Level Sensing}
\author[1]{Berkay~Cetkin}
\author[1]{Lejla~Begic~Fazlic}
\author[1]{Kristof~Ueding}
\author[1]{Rüdiger~Machhamer}
\author[1]{Achim~Guldner}
\author[1]{Lars~Creutz}
\author[1]{Stefan~Naumann}
\author[1]{Guido~Dartmann}

\affil[1]{Institute for Software Systems (ISS), Trier University of Applied Sciences, Environmental Campus Birkenfeld, Hoppstädten-Weiersbach, RP 55768 Germany}

\affil[ ]{}
\affil[ ]{\small{Corresponding author: Berkay Cetkin (e-mail: b.cetkin@umwelt-campus.de).

This work was partly funded by the German Federal Ministry of Food and Agriculture (BMEL) project "KI-Pilot" under Grant 2820KI001, the German Federal Ministry for the Environment, Nature Conservation, Nuclear Safety, and Consumer Protection (BMUV) project "KIRA" under Grant 67KI32013B, and the German Federal Ministry for Economic Affairs and Climate Action (BMWK) project "EASY" under Grant 01MD22002D. 

Patent application (File No. 10 2024 109 324.2) related to the methodology discussed in this paper was filed with the German Patent and Trademark Office on April 3, 2024.

Resources from our research are accessible at \url{https://gitlab.rlp.net/rgdsai/sirec/}.}}
\date{}

\maketitle

\begin{abstract}
In this paper, we propose an innovative method for determining the fill level of containers, such as trash cans, addressing a critical aspect of waste management. The method combines spatial impulse response analysis with machine learning (ML) techniques, offering a unique and effective approach for sound-based classification that can be extended to various domains beyond waste management. By employing a buzzer-generated sine sweep signal, we create a distinctive signature specific to the fill level of the waste container. This signature, once accurately decoded, is then interpreted by a specially developed ensemble learning algorithm. Our approach achieves a classification accuracy of over 90\% when implemented locally on a development board, optimizing operational efficiencies and eliminating the need to delegate complex classification tasks to external entities. Using low-cost and energy-efficient hardware components, our method offers a cost-effective approach that contributes to sustainable and efficient waste management practices, providing a reliable and locally deployable solution.
\end{abstract}


\section{Introduction}
\label{sec:introduction}
The domain of technological innovation continually seeks solutions that transcend conventional boundaries and offer versatility across diverse domains. This situation creates opportunities for inventive solutions and the uptake of new technologies and methods, not only in waste reduction and transformation but also in addressing broader sustainability challenges.
In the domain of fill management, the application of artificial intelligence (AI) is revolutionizing traditional practices, offering smarter, more efficient solutions. Through AI techniques like machine learning (ML) and deep learning (DL), systems can now forecast patterns of waste generation, streamline collection routes, and improve recycling processes~\cite{Fang2023,Sinthiya2022}. This integration of technology not only improves operational efficiency but also contributes significantly to environmental sustainability. One of the major challenges is improving the efficiency of recycling processes. This includes better sorting of recyclables, reducing contamination in recycling streams, and developing more effective methods for recycling various materials~\cite{Sakai}.

In this study, we introduce a novel sound-based classification approach for measuring the fill levels of waste bins using spatial impulse response analysis and ML methods. This cost-effective technique relies on affordable, energy-efficient hardware and aims to optimize waste collection processes, ultimately contributing to more efficient and sustainable waste management practices. Our methodology can be applied to a wide range of application scenarios, as it enables the execution of classification tasks using sound.

The paper is set up as follows: \Cref{sec:introduction} serves as a comprehensive foundation, beginning with the definition of the fill management problem and the establishment of the study's context, followed by a review of relevant literature and various approaches within the field. It then transitions into a discussion of our research methodologies, encompassing a range of techniques and methods employed in our investigation.
In \Cref{sec:SystemModel}, we detail our proposed system's design and architecture. This section also introduces our Sonic Impulse Response Ensemble Classifier (SIREC) algorithm and describes its concept and operational details. \Cref{sec:Validation} is dedicated to presenting our findings, where we validate our results through various scenarios. The paper concludes in \Cref{sec:Conclusion} with a discussion that synthesizes our main insights and reflects on the study's implications.

\noindent Our main contribution enables us to:
\begin{enumerate}
   \item Introduce an innovative approach to acoustic sound classification, tailored for a diverse array of application scenarios. Our innovation lies in the development of a novel AI algorithm, crafted in-house. This algorithm is a practical solution designed for real-world applications like level detection and sound localization, demonstrating versatility and adaptability.
   \item Achieve local sound classification with reliable accuracy, on an ESP8266 microcontroller. This aspect of our work is particularly noteworthy as it merges solid performance with practical applicability, ensuring our methodology is both efficient and accessible.
   \item Rely exclusively on cost and energy effective hardware throughout the entire process. This strategic decision significantly broadens the potential for widespread application, democratizing access to advanced acoustic sound classification technology. Our method makes sound processing affordable and ensures high quality. This opens up new ways to use AI in sound processing.
\end{enumerate}

\subsection{Related Work}

In recent years, innovative methods, particularly those focusing on ML techniques, have been investigated in the context of waste management. ML algorithms play a significant role in various stages of waste management, including waste generation, collection, and transportation. These technologies are also employed in different waste disposal methods such as composting and incineration \cite{Xia_2021}. Recently, researchers referenced in Fang et al. \cite{Fang2023} have conducted an analysis of different ML strategies used in handling waste management. This study focuses on the improvement of processes such as the generation, detection, collection, and sorting of municipal solid waste. The study further highlights how AI models and ML algorithms are applied in scenarios involving the prediction, and recycling of solid waste. Additionally, in studies presented in \cite{RAHMAN2022,Saleem2019,yelekar2023iotbased}, the authors propose a model that combines DL with the internet of things (IoT) to enhance classification and real-time data monitoring in fill level management. Another innovative approach is the Smart Ensembled Framework for Waste Management (SEFWaM), introduced by authors in \cite{Goel2023}. SEFWaM combines DL and computer vision to categorize garbage into various classes, thereby improving waste management processes. In a different study \cite{Srizon2023}, authors discuss an ensemble architecture for waste type recognition using transfer learning techniques and a custom lightweight convolutional neural network (CNN). Meanwhile, authors in \cite{ADEDEJI2019} put forward an intelligent waste classification system that employs a combination of CNN for feature extraction and support vector machine (SVM) for classification. The results demonstrate high accuracy in waste categorization using this approach. A study by \cite{HOY2022} discusses the use of Bayesian-optimized neural networks and ensemble learning to predict different types of city waste. It emphasizes how separating waste and leveraging technology can enhance waste management sustainability. While some authors \cite{AlMashhadani2023, Bai2018,AlFoudery2018TrashBS,Samann_2017, SOHAG2020} proposed IoT-based waste management systems, they did not provide a structural design based on DL or ML techniques. In a recent study \cite{Hassanein2023}, the authors present an intelligent waste management system (IWMS) for smart cities using IoT technology. Their focus is on optimizing energy consumption in smart waste bins, handling missing data with a k-nearest neighbors (k-NN) algorithm, and improving the efficiency of waste truck routes. Authors in \cite{Machhamer2019} present the development of a novel indoor positioning system that utilizes online learning algorithms for sound-based localization in cyber-physical environments, demonstrating significant improvements in adaptability and performance.

\subsection{System Overview}
\begin{figure}[!t]
 \centering
\includegraphics[width=\columnwidth]{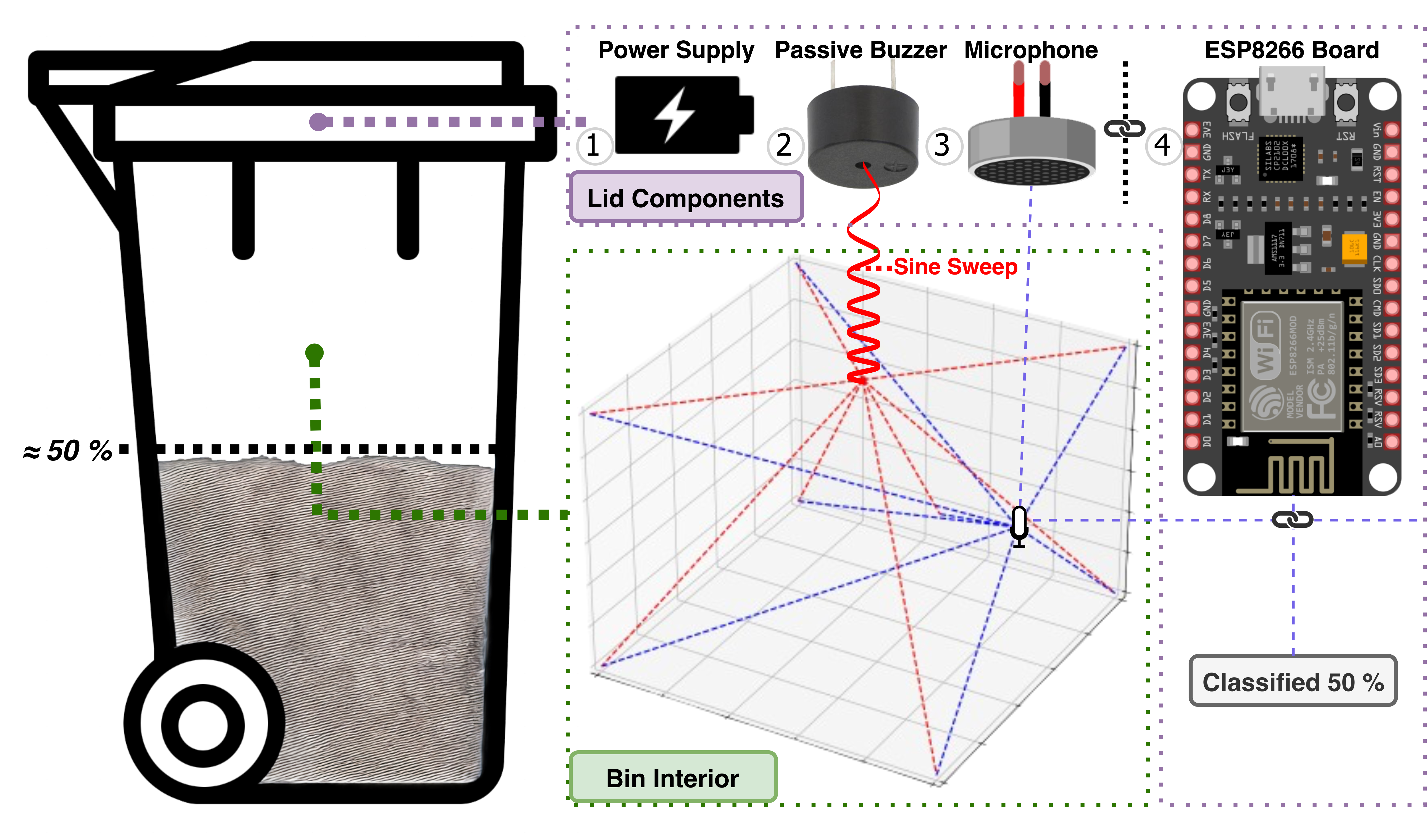}
  \caption{Smart waste bin with integrated level sensing.}
  \label{fig:SmartBin}
\end{figure}

This project focuses on the development of an IWMS that utilizes on-device AI for the classification of waste container fill levels. As a proof of concept, the inner side of a waste container's lid was equipped with an ESP8266, a compact Wi-Fi-capable micro-controller \cite{esp8266}, to which a passive buzzer and a microphone were attached, as shown in \Cref{fig:SmartBin}. The ESP8266 board (\textit{4}) served as the central processing unit of the system, generating sound signals (\textit{sine sweeps}) through the passive buzzer (\textit{2}) and capturing the resulting acoustic response (\textit{bin interior}) via the microphone (\textit{3}). The captured and preprocessed (see \Cref{sec:SystemModel}) signal was then transmitted to a Raspberry Pi for storage using MQTT \cite{mqtt}, a lightweight messaging protocol designed for small sensors and mobile devices.

The primary objective of this research is to accurately classify the fill levels of waste containers into five distinct categories: empty (0\%), quarter-filled (25\%), half-filled (50\%), three-quarters filled (75\%), and completely filled (100\%). To simulate waste materials, straw and cardboard were chosen due to their varying acoustic response characteristics. Central to our research is the SIREC algorithm, which is based on ensemble learning principles. Ensemble learning algorithms are well-established in the field of ML and are known for their ability to generate robust and accurate predictions by combining the decisions of multiple models. In this study, we employ an ensemble of decision tree (DT) models to classify the fill levels. DTs were selected for their versatility in handling complex datasets and their ability to provide transparent and interpretable decision pathways.

\subsection{Spectral Subtraction}
Spectral subtraction \cite{Vaseghi1996,ch12} is a noise reduction technique that leverages the principles of Fourier transform that decomposes a function into its constituent frequencies. Specifically, it employs the discrete Fourier transform (DFT), which is the application of Fourier analysis to a sequence of discrete values. DFT converts a time-discrete signal \(x[n]\) of length \(N\) into its frequency domain representation \(X[k]\), also comprising \(N\) values. The formal definition of DFT and inverse discrete Fourier transform (IDFT) is described in \cite{bracewell1978fourier}.

In the context of DFT, the fast Fourier transform (FFT) \cite{cooley1965algorithm} stands out as particularly noteworthy. FFT is a highly efficient algorithm for computing the DFT and its inverse, significantly optimizing the computational time required for these calculations, thus making it indispensable in various fields, including digital signal processing. Spectral subtraction aims to restore the power spectrum (the squares of the absolute values of the Fourier transform) or the magnitude spectrum (the absolute values of the Fourier transform) of a signal observed in additive noise by subtracting an estimate of the noise spectrum from the noisy signal spectrum \cite {Vaseghi1996}. This noise spectrum is typically estimated during periods when the signal is absent, and only noise is present, under the assumption that noise is a stationary or slowly varying process and that its spectrum does not significantly change during these periods. This premise, particularly advantageous for short-duration recordings, along with the relatively low computational complexity of spectral subtraction that can be managed by simple micro-controllers, contributed significantly to the selection of this method for our study.

\subsection{Spatial Impulse Response}
The impulse response is an essential construct in signal processing, representing the output of a linear time-invariant (LTI) system when subjected to an impulse \cite{lathi2010signal}. An impulse in signal processing is characterized as a transient signal of short duration. For continuous systems, this is often described by the Dirac delta function\cite{bracewell1978fourier}, while discrete systems utilize the Kronecker delta function\cite{bracewell1978fourier}.

LTI systems are characterized by linearity, adhering to the principle of superposition, and time-invariance, maintaining consistent responses over time shifts, as detailed in \cite{lathi2010signal}. The convolution process, integral to LTI systems, involves the combination of an impulse response with an input signal to yield the output signal. Building upon the concept of impulse response, the spatial impulse response—also known as the room impulse response (RIR) \cite{Klein:794700}—describes the temporal sequence of sound reflections and reverberation in a room triggered by a short acoustic pulse. The RIR is the impulse response of a room to a sound impulse, encapsulating all pertinent information about the acoustic properties of the space and enabling the analysis of audibility and room acoustics. In this work, the recorded data comprise RIRs that are specifically utilized as the foundational dataset for training the ML model employed in our analysis.

\section{System Model}
\label{sec:SystemModel}
\begin{figure*}
 \center
  \includegraphics[width=\textwidth]{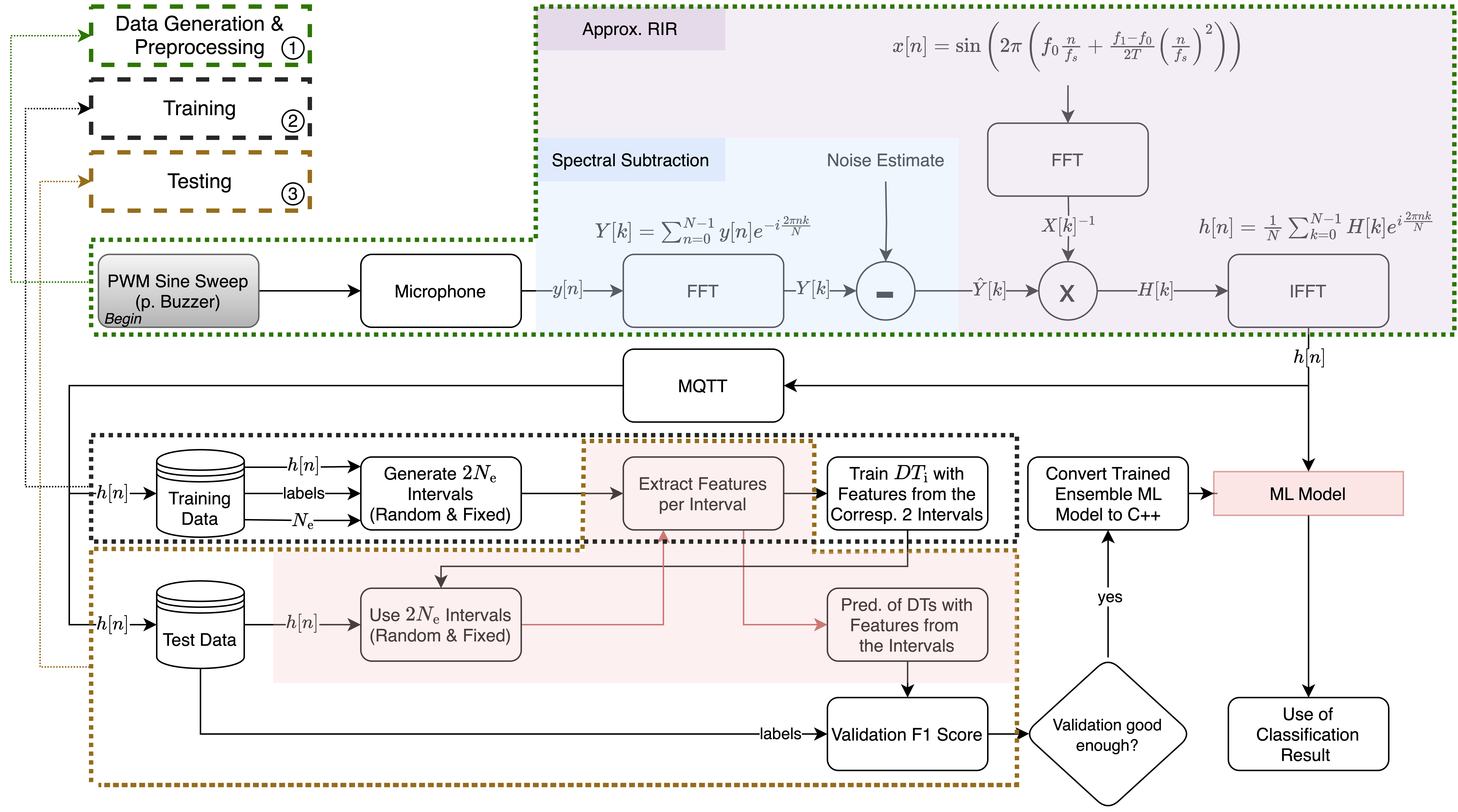}
  \caption{Flow chart representation of our system's methodological approach.}
  \label{fig:Flowchart}
\end{figure*}

The concept of this study is depicted in \Cref{fig:Flowchart}. As shown in the figure, the flowchart consists of three main areas: (\textit{1}) Data Generation \& Preprocessing, (\textit{2}) Training, and (\textit{3}) Testing. In the following subsections, these areas will be explained in more detail.

\subsection{Data Generation \& Preprocessing}
\begin{figure}[!ht]
 \centering
\includegraphics[width=\columnwidth]{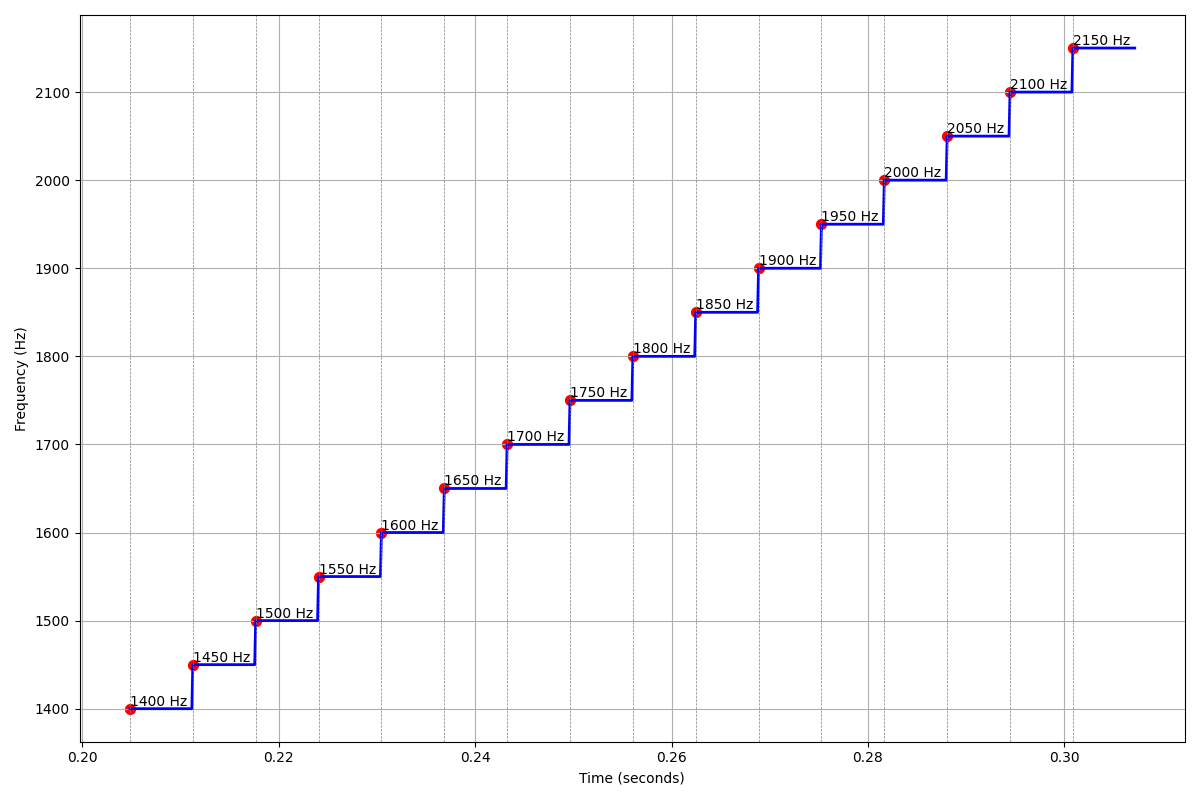}
  \caption{Employed sine sweep signal.}
  \label{fig:SineSweep}
\end{figure}

Starting with the gray area in the presented flowchart, an ESP8266 utilizes pulse width modulation (PWM)--a method used to encode the amplitude of a signal into a duty cycle, which controls the power delivered to a load--to generate a linear sinusoidal sweep. This sweep serves as an auditory input signal produced via a passive buzzer. The signal is subsequently captured by a microphone, representing the output signal \( y[n] \).
Spectral subtraction is then employed for noise reduction of the output signal (see pseudo-code \Cref{alg:SpecSub}). This involves converting the signal into its frequency spectrum via FFT, followed by subtracting the estimated noise. This noise estimate is derived from a designated section recorded by the microphone where the sinus sweep is inaudible, hence, only background noise is present. The noise is transformed into its frequency spectrum, and then the disruptive frequencies are subtracted from the output signal \( y[n] \). Post spectral subtraction, the RIR is approximated (see pseudo-code \Cref{alg:RIR}). Given that an impulse response uniquely identifies an LTI system, the most straightforward choice for a measuring signal to determine the RIR would be a Dirac or Kronecker impulse. An idealized impulse, such as the Dirac delta function in continuous systems, theoretically spans the entire frequency range, signifying that all frequencies are stimulated simultaneously and with equal intensity, as evidenced by its Fourier transform yielding a constant value across all frequencies \cite{Riley_Hobson_Bence_2006}. Similarly, the Kronecker delta function \cite{bracewell1978fourier} in discrete systems exhibits an analogous property, with its Fourier transform resulting in a constant value for all frequency indices \( k \), limited to a range up to the so-called Nyquist frequency\cite{lathi2010signal}. However, generating such idealized impulses with a sound source is impractical, for instance, due to the infinitely short duration of the signals\cite{oppenheim2010signals}. Hence, alternative methods that mimic the properties of the ideal impulses but are feasible in reality are often used in electro-acoustic systems to stimulate sound impulses. The linear sinusoidal sweep is one such implementable signal\cite{Klein:794700}. It uses an excitation signal that covers as broad a frequency range as possible to enable a comprehensive measurement of a room's acoustic properties. \Cref{eq:1} describes such a linear sinusoidal sweep that varies the frequency from \( f_0 \) to \( f_1 \) over time \( T \), where \( f_s \) denotes the sampling rate \cite{farina2000simultaneous}.
\begin{equation} 
x[n] = \sin\left(2\pi\left(\frac{f_0 n}{f_\text{s}} + \frac{f_1 - f_0}{2T}\left(\frac{n}{f_\text{s}}\right)^2\right)\right)
\label{eq:1}
\end{equation}
\Cref{fig:SineSweep} illustrates the actual signal employed in this work. It is important to note that, although this signal appears as a step function due to the discrete increments of frequency, it still maintains the key characteristics of a linear sinusoidal sweep, specifically through uniform frequency increments at consistent time intervals. This particular implementation was chosen based on empirical findings; it demonstrated superior signal clarity and significantly reduced distortions when generated through a cost-effective passive buzzer using PWM signals. When considering the mathematical limit as the duration of each frequency approaches zero, the function aligns more closely with a true linear model. However, practical constraints necessitate a balance between idealized models and achievable outcomes. Thus, although the signal does not represent a perfect linear sweep, its operational performance fulfills the criteria for a linear sweep, especially within the context of our objectives—namely, the comprehensive measurement of a room's acoustic properties.

\begin{algorithm}[!b]
\caption{Spectral Subtraction - Arduino}
\label{alg:SpecSub}
\begin{algorithmic}[1]
\State $\text y \gets \text{record}(f_{0},f_{1},f_\text{step})$
\State $\alpha \gets 1.0$
\State $\text{noiseSample}_\text{end} \gets \text{startSampleBuzzer} - 200$
\State $\text{noiseSample}_\text{start} \gets \text{noiseSample}_\text{end} - 2^{10}$

\For{$i \gets \text{noiseSample}_\text{start},..., \text{noiseSample}_\text{end}$}
    \State $\text n[i] \gets \text y[i]$
\EndFor
  
\For{$i \gets 0,...,N-1$}
    \State $\text y_\text{windowed}[i] \gets \text y[i] \cdot \text w[i]$
    \State $\text n_\text{windowed}[i] \gets \text n[i] \cdot \text w[i]$
\EndFor

\State $\text Y_\text{windowed} \gets \text{FFT}(\text y_\text{windowed})$
\State $|\text{c}_\text{y}| \gets |\text Y_\text{windowed}|$
\State $\text N_\text{windowed} \gets \text{FFT}(\text n_\text{windowed})$
\State $|\text{c}_\text{n}| \gets |\text N_\text{windowed}|$

\For{$i \gets 0,...,N-1$}
    \State $\text{phase}_\text{y}[i] \gets \text{atan2}(\text Y_\text{windowedIm}[i], \text Y_\text{windowedRe}[i])$
    \State $\text Y_\text{rightMag}[i] \gets \max(0.0, |\text{c}_\text{y}[i]| - \alpha \cdot |\text{c}_\text{n}[i]|)$
\EndFor

\State $\text y \gets \text{IFFT}(\text Y_\text{rightMag} \cdot \cos(\text{phase}_\text{y}), \text Y_\text{rightMag} \cdot \sin(\text{phase}_\text{y}))$
\end{algorithmic}
\end{algorithm}

\begin{algorithm}[!b]
\caption{Approx. RIR - Arduino}
\label{alg:RIR}
\begin{algorithmic}[1]
\State $\text x \gets \text{calcLinearSineSweep}(f_{0},f_{1},f_\text{step})$
\State $\text y \gets \text{record}(f_{0},f_{1},f_\text{step})$

\For{$i \gets 0,...,N-1$}
    \State $\text x_\text{windowed}[i] \gets \text x[i] \cdot \text w[i]$
    \State $\text y_\text{windowed}[i] \gets \text y[i] \cdot \text w[i]$
\EndFor

\State $\text X_\text{windowed} \gets \text{FFT}(\text x_\text{windowed})$
\State $|\text c_\text{x}| \gets |\text X_\text{windowed}|$
\State $\text Y_\text{windowed} \gets \text{FFT}(\text y_\text{windowed})$
\State $|\text c_\text{y}| \gets |\text Y_\text{windowed}|$

\For{$i \gets 0,...,N-1$}
    \State $\text{phase}_\text{y}[i] \gets \text{atan2}(\text Y_\text{windowedIm}[i], \text Y_\text{windowedRe}[i])$
    \State $\text{rir}_\text{mag}[i] \gets \max(0.0, |\text c_\text{y}[i]| / |\text c_\text{x}[i]|)$
\EndFor

\State $\text{rir} \gets \text{IFFT}(\text{rir}_\text{mag} \cdot \cos(\text{phase}_\text{y}), \text{rir}_\text{mag} \cdot \sin(\text{phase}_\text{y}))$
\end{algorithmic}
\end{algorithm}

To extract the RIR from the recorded sinusoidal sweep signal, an inversion technique is applied, leveraging the known excitation signal to approximate the room's response. Essentially, the recorded signal is "unfolded" with the original sweep signal to obtain the RIR. This unfolding, also referred to as deconvolution, is the inverse of the convolution operation, attempting to deduce an unknown function \( h \) from the convolution result \( y \) of two functions \( x \) and \( h \), given \( x \) and \( y \) \cite{Klein:794700}. The convolution operation is always computable, whereas its inverse, deconvolution, may not always be feasible due to potential information loss\cite{Klein:794700}. An approximate solution for determining the RIR is to utilize the frequency domain, based on the principle that the Fourier transform of a convolution of two functions is the product of their individual Fourier transforms\cite{Skinner}. Consequently, the (approximate) RIR is the inverse Fourier transform of the division of the Fourier transform of the recorded sinusoidal sweep (output signal, \( \hat{Y}[k] \)) by the Fourier transform of the sinusoidal sweep (input signal, \( X[k] \)).

\subsection{Training \& Testing}
Following the extraction of RIRs, the data is transmitted to a Raspberry Pi via MQTT. There, Python scripts manage and utilize the data for two primary processes, depicted as numbers \textit{2} and \textit{3} in \Cref{fig:Flowchart}, which involve the SIREC algorithm developed for this project. During the training process, RIRs corresponding to varying fill levels of the waste container are collected and labeled. The collection of the RIRs can be represented by Equation~\eqref{eq:2}: 

\begin{equation}
\mathbf{X}_{S,1:N} = \begin{bmatrix}
v_{1,1} & v_{1,2} & \cdots & v_{1,p_i} 
& \cdots & v_{1,q_i} & \cdots & v_{1,N} \\
v_{2,1} & v_{2,2} & \cdots & v_{2,p_i} 
& \cdots & v_{2,q_i} & \cdots & v_{2,N} \\
\vdots & \vdots & \ddots & \vdots  & \ddots & \vdots  & \ddots & \vdots \\
v_{S,1} & v_{S,2} & \cdots & v_{S,p_i} 
& \cdots & v_{S,q_i} & \cdots & v_{S,N}
\end{bmatrix}
\label{eq:2}
\end{equation}
The corresponding labels are given by Equation~\eqref{eq:3}:
\begin{equation}
    \mathbf{y} = 
\begin{bmatrix} l_{1}, & l_{2}, & \ldots, & l_{S} \end{bmatrix}^T
\label{eq:3}
\end{equation}
where \(\mathbf{X}_{S,1:N} \in \mathbb{R}^{S \times N}\), with $S$ as the number of samples (\(S \in \mathbb{N} \text{ with } S > 1\)) and $N$ as the length of the RIRs (\(N \in \mathbb{N} \text{ with } N > 1\)), and \(\mathbf{y} \in \mathbb{L}^{S} \text{ where } \mathbb{L} \text{ is the set of possible labels}\). 
An ensemble ML model, specifically the SIREC model, is then trained using these RIRs, employing a predetermined number of estimators \(N_{e} \in \mathbb{N}_1\), which are DTs. For each tree, two intervals are generated: a random interval, whose length varies within a adjustable maximum and minimum range across the total length of the RIRs, and a fixed interval, matching the length of the random interval but positioned at the start of the RIRs. Each random interval can be represented as a submatrix of \(\mathbf{X}_{S,1:N}\) as follows in Equation \eqref{eq:4}: 
\begin{equation} 
\mathbf{X}_{S,p_i:q_i} = \begin{bmatrix}
v_{1,p_i} & v_{1,p_i+1} & \ldots & v_{1,q_i} \\
v_{2,p_i} & v_{2,p_i+1} & \ldots & v_{2,q_i} \\
\vdots  & \vdots  & \ddots & \vdots \\
v_{S,p_i} & v_{S,p_i+1} & \ldots & v_{S,q_i}
\end{bmatrix}
\label{eq:4}
\end{equation}
where \(\mathbf{X}_{S,p_i:q_i} \in \mathbb{R}^{S \times (q_i - p_i + 1)}\) with \(p_i\) and \(q_i\) as the start and end column indices of \(\mathbf{X}_{S,1:N}\) for the \(i\)-th random interval (\(i \in \{1, 2, \ldots, N_e\}\)), and \(p_i, q_i \in \{1,2,\ldots,N\} \text{ with } p_i < q_i \).
For the random intervals, two feature classes are extracted. Let \(\mathbf{\hat{v}}_{s_{i}}\) be defined by Equation \eqref{eq:5} as follows:

\begin{equation}
\mathbf{\hat{v}}_{s_{i}} = \begin{bmatrix} v_{s, p_i} & v_{s, p_i + 1} & \ldots & v_{s, q_i} \end{bmatrix}^T 
\label{eq:5}
\end{equation}

where \(\mathbf{\hat{v}}_{s_{i}}\) is the \(s\)-th row of \(\mathbf{X}_{S,p_i:q_i}\). Then, \(\forall s = 1, 2, \ldots, S: \forall i = 1, 2, \ldots, N_{e}\),  the following transformations are applied as shown in Equations~\eqref{eq:6} and \eqref{eq:7}:
\begin{equation}
    \mathbf{\hat{x}}_{s_{i}} = \left| \text{FFT}\left
( \mathbf{\hat{v}}_{s_{i}} \right) \right| 
\label{eq:6}
\end{equation}
\begin{equation}
\hat{z}_{s_{i}} = \frac{\min(\mathbf{\hat{v}}_{s_{i}})}{\max(\mathbf{\hat{v}}_{s_{i}})}
\label{eq:7}
\end{equation}
where \(\mathbf{\hat{v}}_{s_{i}} \in \mathbb{R}^{(q_i-p_i+1)}\), \(\mathbf{\hat{x}}_{s_{i}}\) are the FFT features from \(\mathbf{\hat{v}}_{s_{i}}\) (the magnitude of the frequency spectrum) with \(\mathbf{\hat{x}}_{s_{i}} \in \mathbb{R}^{2^{\lfloor log_2(q_i - p_i) + 0.5\rfloor}}\), and \(\hat{z}_{s_{i}}\) is the min/max ratio feature from \(\mathbf{\hat{v}}_{s_{i}}\) with \(\hat{z}_{s_{i}} \in \mathbb{R}\). Each fixed interval can also be represented as a submatrix of \(\mathbf{X}_{S,1:N}\): 
\begin{equation}
\mathbf{X}_{S,1:q_i-p_i+1} = \begin{bmatrix}
v_{1,1} & v_{1,2} & \ldots & v_{1,q_i-p_i+1} \\
v_{2,1} & v_{2,2} & \ldots & v_{2,q_i-p_i+1} \\
\vdots  & \vdots  & \ddots & \vdots \\
v_{S,1} & v_{S,2} & \ldots & v_{S,q_i-p_i+1}
\end{bmatrix}
\end{equation}
where \(\mathbf{X}_{S,1:q_i-p_i+1} \in \mathbb{R}^{S \times (q_i - p_i + 1)}\). 

For the fixed intervals, two feature classes are also extracted.
Let \(\mathbf{v}_{s_{i}}\) be defined by Equation~\eqref{eq:8} as follows:

\begin{equation}
\mathbf{v}_{s_{i}} = [v_{s, 1}, v_{s, 2}, \ldots, v_{s, q_i-p_i+1}]^T
\label{eq:8}
\end{equation}

where \(\mathbf{v}_{s_{i}}\) is the \(s\)-th row of \(\mathbf{X}_{S,1:q_i-p_i+1}\).

 Then, \(\forall s = 1, 2, \ldots, S: \forall i = 1, 2, \ldots, N_{e}:\)
 \begin{equation}
 \label{eq:meanDiff}
\mu_{s_{i}} = \frac{\sum_{k=1}^{q_i-p_i} (v_{s,k} - v_{s,k+1})}{q_i-p_i}
 \end{equation}
 \begin{equation}
  \label{eq:stdDiff}
\sigma_{s_{i}} = \sqrt{
\frac{1}{q_i-p_i} \sum_{k=1}^
{q_i-p_i} \left(
v_{s,k} - v_{s,k+1} - \mu_{s_{i}}\right)^2}
 \end{equation}
 where \(\mathbf{v}_{s_{i}} \in \mathbb{R}^{(q_i-p_i+1)}\), \(\mu_{s_{i}}\) is a specific mean feature from \(\mathbf{v}_{s_{i}}\) with \(\mu_{s_{i}} \in \mathbb{R}\), and \(\sigma_{s_{i}}\) is a specific standard deviation feature from  \(\mathbf{v}_{s_{i}}\) with \(\sigma_{s_{i}} \in \mathbb{R}\).

\begin{figure}[!b]
 \centering
\includegraphics[width=\columnwidth]{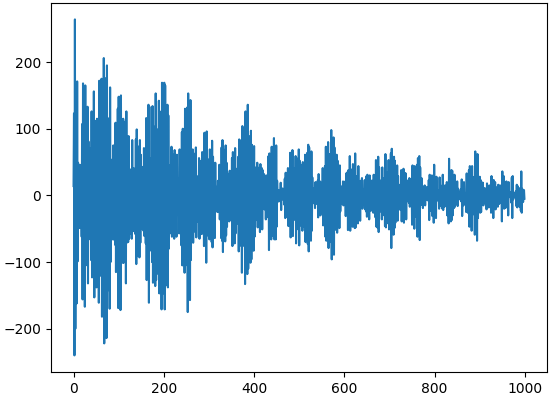}
  \caption{Example of a calculated RIR.}
  \label{fig:RIR}
\end{figure}

For each DT and the associated two intervals (random \& fixed), the four feature classes outlined here are extracted to train the respective tree and infer during the prediction process. The two feature classes for the fixed interval are based on novel approaches and have the following rationale: Time series data can have varying structures, and in some cases, they may exhibit symmetry along the x-axis without a vertical shift. In this project, the RIR is calculated in a manner that closely exhibits these two properties. An example of an extracted RIR is shown in \Cref{fig:RIR}. Considering this structure of the RIRs, calculating the mean would yield a value close to zero for different classes, making it also problematic for the standard deviation, which depends on the mean. In light of this, we propose alternative calculation methods (\Cref{eq:meanDiff} \& \Cref{eq:stdDiff}), computing the difference between each pair of adjacent values across all values of a time series variable to determine the mean and standard deviation of the derived series, which is one value shorter than the original. This approach captures not only the magnitude of the mean and standard deviation but also the characteristic property of each RIR, combined with these two measures.

\begin{figure*}
 \center
  \includegraphics[width=\textwidth]{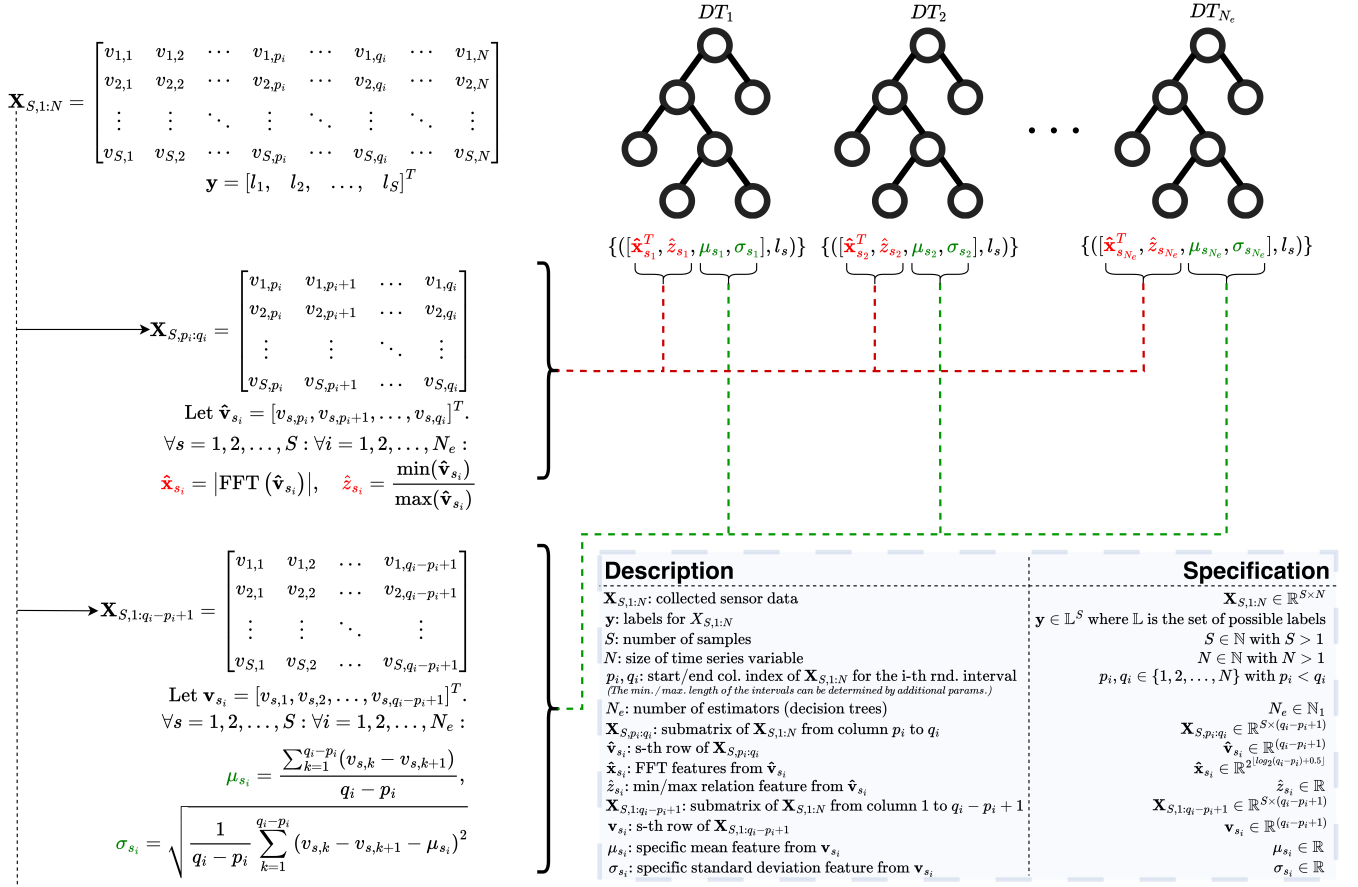}
  \caption{Sonic Impulse Response Ensemble Classifier (SIREC) methodology.}
  \label{fig:SIRECMethodology}
\end{figure*}

Each of the \(N_\text{e}\) trees is trained using this feature extraction methodology, forming an independent ML model that, in combination, constitutes SIREC, the complete ensemble ML model (\Cref{fig:SIRECMethodology}). In the testing process, depicted as number \textit{3} in \Cref{fig:Flowchart}, the model's accuracy is validated using a portion of the collected RIRs and their corresponding fill level labels. Here, SIREC utilizes the intervals created during its training process, extracts relevant features, and aggregates predictions from all its DTs to derive a final result. The F1 score metric is then used to determine if the validation is sufficiently accurate to, if successful, convert the trained and validated SIREC model into C++ code. This code can then be seamlessly uploaded to the ESP8266 board using the Arduino IDE \cite{arduino}--a development environment that facilitates programming and interfacing with various micro-controllers and modules. Thus, enabling local predictions using the trained ML model without the need to delegate the classification task to external resources.

\section{Validation}
\label{sec:Validation}
A series of experiments were conducted to validate our methodology, particularly the SIREC algorithm, by setting up distinct test environments tailored to specific variables and aspects. The focus was on evaluating the algorithm's performance and energy efficiency when applied to waste container fill level classification. The validation process also explored the potential of extending the proposed methodology to other use cases, while discussing validity considerations and challenges to provide a complete assessment of the approach.


\subsection{Performance}
Data was systematically collected under varied conditions to benchmark different classification models. A selection of the characteristics of these data collection environments is presented in \Cref{tab:data}. The first column in the dataset uniquely identifies each entry. The next five columns present the distribution of data across designated categories. The final column indicates the type of filler material used, which is either straw or cardboard (CB). A linear sine-sweep signal (\Cref{fig:SineSweep}), used to derive these measurements, was played between samples 2048 and 3072 of a 4096-sample-long recording captured at a sampling rate of 10kHz. The sample interval prior to this range was utilized for noise estimation, instrumental in noise attenuation via spectral subtraction.

\begin{table}[!t]
\centering
\caption{Extract of Accumulated Measurement Data}
\label{tab:data}
\begin{tabular}{@{}cllllll@{}}
\toprule
\textbf{ID} & \textbf{\begin{tabular}[c]{@{}l@{}}0\textsubscript{\#}\end{tabular}} & \textbf{\begin{tabular}[c]{@{}l@{}}25\textsubscript{\#}\end{tabular}} & \textbf{\begin{tabular}[c]{@{}l@{}}50\textsubscript{\#}\end{tabular}} & \textbf{\begin{tabular}[c]{@{}l@{}}75\textsubscript{\#}\end{tabular}} & \textbf{\begin{tabular}[c]{@{}l@{}}100\textsubscript{\#}\end{tabular}} & \textbf{Mat.} \\ 
\midrule
$T_{1}$ & \begin{tabular}[c]{@{}l@{}}0\%: 20\\ 5\%: 20\\ 10\%: 20\end{tabular} & \begin{tabular}[c]{@{}l@{}}20\%: 20\\ 25\%: 20\\ 30\%: 20\end{tabular} & \begin{tabular}[c]{@{}l@{}}45\%: 20\\ 50\%: 20\\ 55\%: 20\end{tabular} & \begin{tabular}[c]{@{}l@{}}70\%: 20\\ 75\%: 20\\ 80\%: 20\end{tabular} & \begin{tabular}[c]{@{}l@{}}90\%: 20\\ 95\%: 20\\ 100\%: 20\end{tabular} & Straw \\ \midrule
$T_{2}$ & \begin{tabular}[c]{@{}l@{}}0\%: 20\\ 5\%: 20\\ 10\%: 20\end{tabular} & \begin{tabular}[c]{@{}l@{}}20\%: 20\\ 25\%: 20\\ 30\%: 20\end{tabular} & \begin{tabular}[c]{@{}l@{}}45\%: 20\\ 50\%: 20\\ 55\%: 20\end{tabular} & \begin{tabular}[c]{@{}l@{}}70\%: 20\\ 75\%: 20\\ 80\%: 20\end{tabular} & \begin{tabular}[c]{@{}l@{}}90\%: 20\\ 95\%: 20\\ 100\%: 20\end{tabular} & CB \\ \midrule
$T_{3}$ & \begin{tabular}[c]{@{}l@{}}0\%: 10\\ 5\%: 10\\ 10\%: 10\end{tabular} & \begin{tabular}[c]{@{}l@{}}20\%: 10\\ 25\%: 10\\ 30\%: 10\end{tabular} & \begin{tabular}[c]{@{}l@{}}45\%: 10\\ 50\%: 10\\ 55\%: 10\end{tabular} & \begin{tabular}[c]{@{}l@{}}70\%: 10\\ 75\%: 10\\ 80\%: 10\end{tabular} & \begin{tabular}[c]{@{}l@{}}90\%: 10\\ 95\%: 10\\ 100\%: 10\end{tabular} & Straw \\ \midrule
$T_{4}$ & \begin{tabular}[c]{@{}l@{}}0\%: 10\\ 5\%: 10\\ 10\%: 10\end{tabular} & \begin{tabular}[c]{@{}l@{}}20\%: 10\\ 25\%: 10\\ 30\%: 10\end{tabular} & \begin{tabular}[c]{@{}l@{}}45\%: 10\\ 50\%: 10\\ 55\%: 10\end{tabular} & \begin{tabular}[c]{@{}l@{}}70\%: 10\\ 75\%: 10\\ 80\%: 10\end{tabular} & \begin{tabular}[c]{@{}l@{}}90\%: 10\\ 95\%: 10\\ 100\%: 10\end{tabular} & CB \\
\bottomrule
\end{tabular}
\end{table}

\begin{figure*}
 \centering
\includegraphics[width=\textwidth]{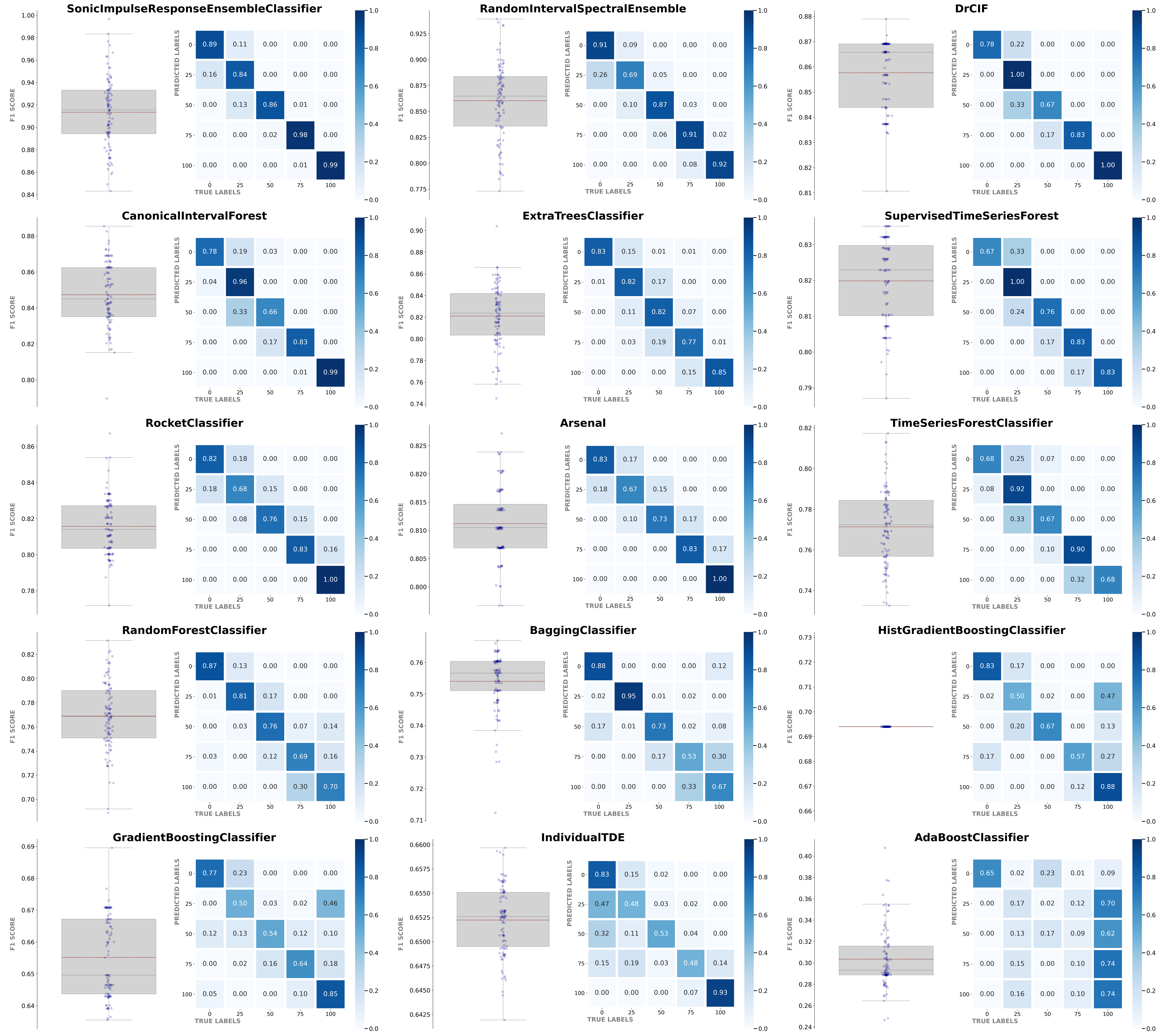}
  \caption{Box plots (F1 score; red dashed line indicating the mean; solid line indicating the median) and confusion matrices (normalized with respect to true labels) for the considered ML models over 100 iterations of training and testing. Training data: $T_{1}$ \& $T_{2}$, Test data: $T_{3}$ \& $T_{4}$. Sample segment length: 300.}
  \label{fig:F1Results}
\end{figure*}

\Cref{fig:F1Results} displays the classification accuracy results of various ML models, represented through box plots for the F1 score metric and confusion matrices normalized with respect to the true labels. The results are based on 100 iterations of training and testing, considering a segment length of 300. It is important to note that, throughout this paper, the term "segment length" refers to the length of a subsection extracted from the total duration of the RIRs. The algorithms presented in \Cref{fig:F1Results}, with the exception of SIREC which has been previously discussed, are selected from the sktime \cite{sktime} and sklearn \cite{scikit-learn} libraries, which are widely used in ML for time series and general-purpose classification tasks, respectively. Notably, all ensemble-based algorithms in this comparison were configured with an equal number of estimators (100), keeping all other parameters at their default values to ensure a consistent basis for comparison.

\begin{figure}[!t]
 \centering
\includegraphics[width=\columnwidth]{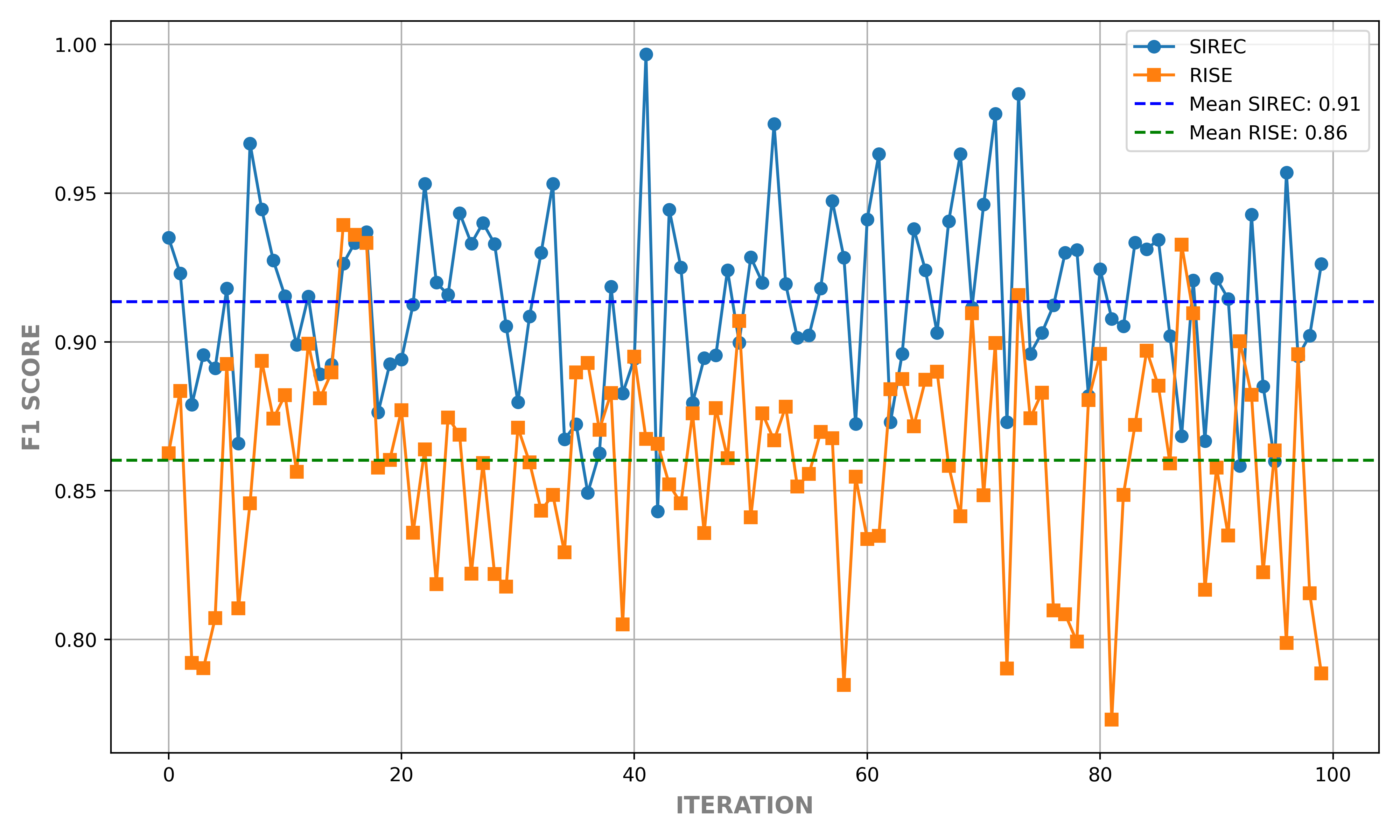}
  \caption{SIREC \& RISE comparison.}
  \label{fig:SIREC_RISE}
\end{figure}

It is particularly noteworthy that SIREC and the Random Interval Spectral Ensemble (RISE) \cite{RISE2019} algorithm, which secured second place in \Cref{fig:F1Results}, share two fundamental commonalities. Both methodologies are predicated on an ensemble of DTs and both engage in the generation of random intervals, within which they extract the magnitude of the frequency spectrum as one of the features. However, there are key advantages that distinguish SIREC from RISE, rendering it more beneficial. Contrary to RISE, SIREC exhibits superior speed and energy efficiency (as will be discussed in the following section and shown in \Cref{tab:data1}), as well as a significant improvement in memory efficiency. This is largely due to the fact that RISE computes the auto-correlation function, with the default setting being 100 auto-correlation terms, which inherently makes SIREC more suited for local operation on micro-controllers for classification tasks. Furthermore, SIREC tends to demonstrate marginally better average accuracy. To support this claim, \Cref{fig:SIREC_RISE}, as an extension to \Cref{fig:F1Results}, provides a detailed depiction of the F1 scores between RISE and SIREC over the 100 iterations.

A significant advantage of the Random Interval Generation concept for feature extraction, as implemented in SIREC, is its inherent reduction or elimination of the need for expert knowledge about the collected data to train the model. For instance, a stochastic search across various randomly generated intervals within specified minimum and maximum lengths allows for the bypassing of such expert knowledge. This stochastic search method ensures the identification of the ideal data segment for feature extraction without necessitating extensive manual data analysis. Moreover, since SIREC is based on DTs, which represent a symbolic form of representation, acquiring expert knowledge about the data post-training is straightforward. Our stochastic search for optimal SIREC parameters (thus, the best random intervals) has achieved accuracies up to 100\% in some cases. The results of this stochastic search, along with the parameters used for SIREC, are presented in \Cref{tab:stochSearchSIREC}.

\begin{table*}[!ht]
\centering
\caption{Stochastic Search for SIREC with Training Data $T_{1}$ \& $T_{2}$, Test Data $T_{3}$ \& $T_{4}$.}
\label{tab:stochSearchSIREC}
\small 
\setlength{\tabcolsep}{5pt} 
\begin{tabular}{cccccc}
\toprule
\textbf{Segment Length} & \textbf{N\textsubscript{e}} & \textbf{Max. Interval Length} & \textbf{Min. Interval Length} & \textbf{Random State} & \textbf{F1 Score} \\
\midrule
460 & 72 & 153 & 17 & 6726550 & 1.00 \\
389 & 171 & 146 & 62 & 26493929 & 0.99 \\
349 & 199 & 189 & 124 & 54072606 & 0.98 \\
382 & 84 & 92 & 49 & 70275997 & 0.97 \\
462 & 156 & 59 & 48 & 61101216 & 0.96 \\
275 & 55 & 154 & 89 & 60576101 & 0.95 \\
\bottomrule
\end{tabular}
\end{table*}

\subsection{Energy Measurements}
In the analysis of energy consumption, our investigation reveals significant insights into the trade-offs between computational energy efficiency and predictive accuracy across three distinct algorithms: our proposed algorithm (SIREC), Time Series Forest Classifier (TSFC)\cite{DENG2013142} and RISE. The measurement method and experiment setup is based upon our previous work \cite{KERN2018199,Guldner2021} and explained in detail by \cite{Guldner2023}. The resource measurements adhere to the methodology and guidelines outlined in the Green Software Measurement Model \cite{GSMM2024}. Measurements were carried out using a computer system that included an Intel Core i5-650 CPU. The system's memory was composed of two 2 GB RAM modules, totaling 4 GB. For storage, the computer utilized a dual-setup, comprising a 500 GB hard disk drive (HDD) and a 250 GB solid-state drive (SSD). The analysis is structured around various scenarios--configurations of segment lengths and numbers of estimators, focusing on key metrics such as energy consumption during training and testing phases, computational time, CPU usage, and F1 score. These scenarios, along with their corresponding results listed in \Cref{tab:data1}, were tested on the same datasets, $T_{1}$ and $T_{2}$ for training, and $T_{3}$ and $T_{4}$ for testing, as detailed in \Cref{tab:data}. It is important to highlight that a random state parameter was employed across all algorithms to ensure comparability and control over experimental variability. A value of 7 was selected arbitrarily for the random state, with no intent to favor any specific algorithm, especially not SIREC.

All algorithms exhibit a pronounced correlation between the configured parameters and their energy consumption, which varies significantly across different settings. Specifically, SIREC's energy consumption during the training phase ranges from 0.06 Wh to 0.47 Wh, and in the testing phase from 0.02 Wh to 0.20 Wh. Despite this variability, SIREC consistently delivers high F1 scores, peaking at 0.94, indicative of its superior accuracy in predictions. This performance comes at the expense of higher energy consumption, especially when compared to TSFC, which maintains remarkable energy efficiency in all configurations and achieves the shortest training and testing times. However, it does so at the cost of lower F1 scores. In contrast, SIREC not only uses more energy but also demands greater CPU resources. This higher CPU usage highlights SIREC’s increased computational requirements, necessary for its superior predictive performance. Meanwhile, TSFC balances energy and CPU efficiency, making it a more suitable option for environments with strict energy constraints, despite its lower F1 scores. RISE presents the most resource-intensive option, with energy consumption and computational time significantly higher than the other algorithms. Despite its high resource demand, RISE achieves competitive F1 scores in certain configurations, suggesting that its resource intensity may be justified in scenarios where predictive accuracy is important.

These observations underscore a fundamental trade-off in algorithm design: while SIREC and RISE require more energy and higher CPU usage to achieve superior or more consistent predictive performance, TSFC prioritizes energy efficiency, which may lead to reduced accuracy. Notably, reducing the segment length and the number of estimators leads to a significant reduction in energy consumption as well as training and testing time, with the latter having a more pronounced impact. This effect is evident across all algorithms, illustrating that while configuration adjustments can lead to greater operational efficiency, they might also affect performance, making a careful choice essential to balance both aspects effectively.

\begin{table*}[!ht]
\centering
\caption{Comparison of Algorithm Performance and Energy Consumption}
\label{tab:data1}
\small 
\setlength{\tabcolsep}{3pt} 
\begin{tabular}{cccccccccc}
\toprule
\multirow{2}{*}{\textbf{Algorithm}} & \multirow{2}{*}{\textbf{Segment Length}} & \multirow{2}{*}{\textbf{N\textsubscript{e}}} & \multirow{2}{*}{\textbf{Mean Power [W]}} & \multicolumn{2}{c}{\textbf{Energy [Wh]}} & \multicolumn{2}{c}{\textbf{Time [s]}} & \multirow{2}{*}{\textbf{CPU Usage [\%]}} & \multirow{2}{*}{\textbf{F1 Score}} \\
 &  &  &  & \textbf{Training} & \textbf{Testing} & \textbf{Training} & \textbf{Testing} &  &  \\
\midrule
\multirow{4}{*}{SIREC} & 300 & 100 & 81.29 & 0.47 & 0.20 & 21.13 & 8.94 & 24.26 & 0.94 \\
 & 300 & 10 & 75.96 & 0.06 & 0.02 & 2.74 & 1.19 & 21.77 & 0.89 \\
 & 200 & 50 & 79.87 & 0.21 & 0.09 & 9.56 & 4.14 & 23.87 & 0.82 \\
 & 100 & 100 & 80.54 & 0.36 & 0.16 & 16.18 & 7.14 & 24.05 & 0.76 \\
\midrule
\multirow{4}{*}{TSFC} & 300 & 100 & 78.60 & 0.13 & 0.03 & 5.78 & 1.33 & 23.10 & 0.78 \\
 & 300 & 10 & 65.09 & 0.02 & 0.00 & 1.05 & 0.37 & 17.37 & 0.77 \\
 & 200 & 50 & 73.29 & 0.06 & 0.00 & 2.80 & 0.63 & 19.95 & 0.76 \\
 & 100 & 100 & 76.44 & 0.08 & 0.02 & 3.72 & 0.69 & 22.23 & 0.65 \\
\midrule
\multirow{4}{*}{RISE} & 300 & 100 & 82.63 & 8.70 & 4.15 & 376.05 & 184.64 & 24.64 & 0.87 \\
 & 300 & 10 & 80.45 & 0.88 & 0.43 & 39.37 & 19.43 & 24.41 & 0.78 \\
 & 200 & 50 & 81.73 & 3.86 & 1.80 & 167.63 & 82.40 & 24.60 & 0.78 \\
 & 100 & 100 & 82.16 & 5.13 & 2.41 & 221.80 & 109.10 & 24.62 & 0.73 \\
\bottomrule
\end{tabular}
\end{table*}

\subsection{Extended Use Cases}
\begin{figure}[!b]
 \centering
\includegraphics[width=\columnwidth]{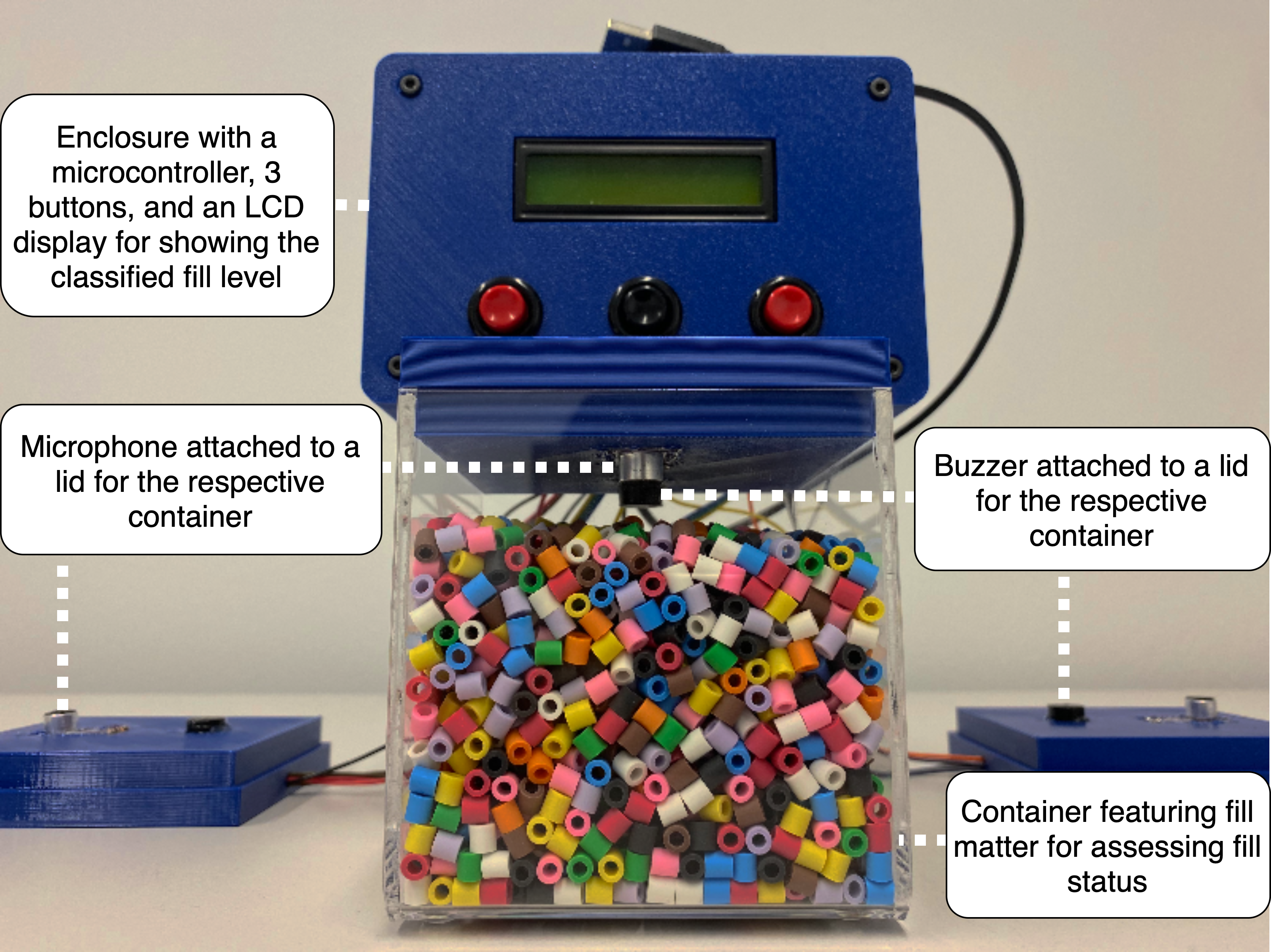}
  \caption{Fill level demonstrator.}
  \label{fig:FillLevelDemonstrator}
\end{figure}

\begin{figure}[!b]
 \centering
\includegraphics[width=\columnwidth]{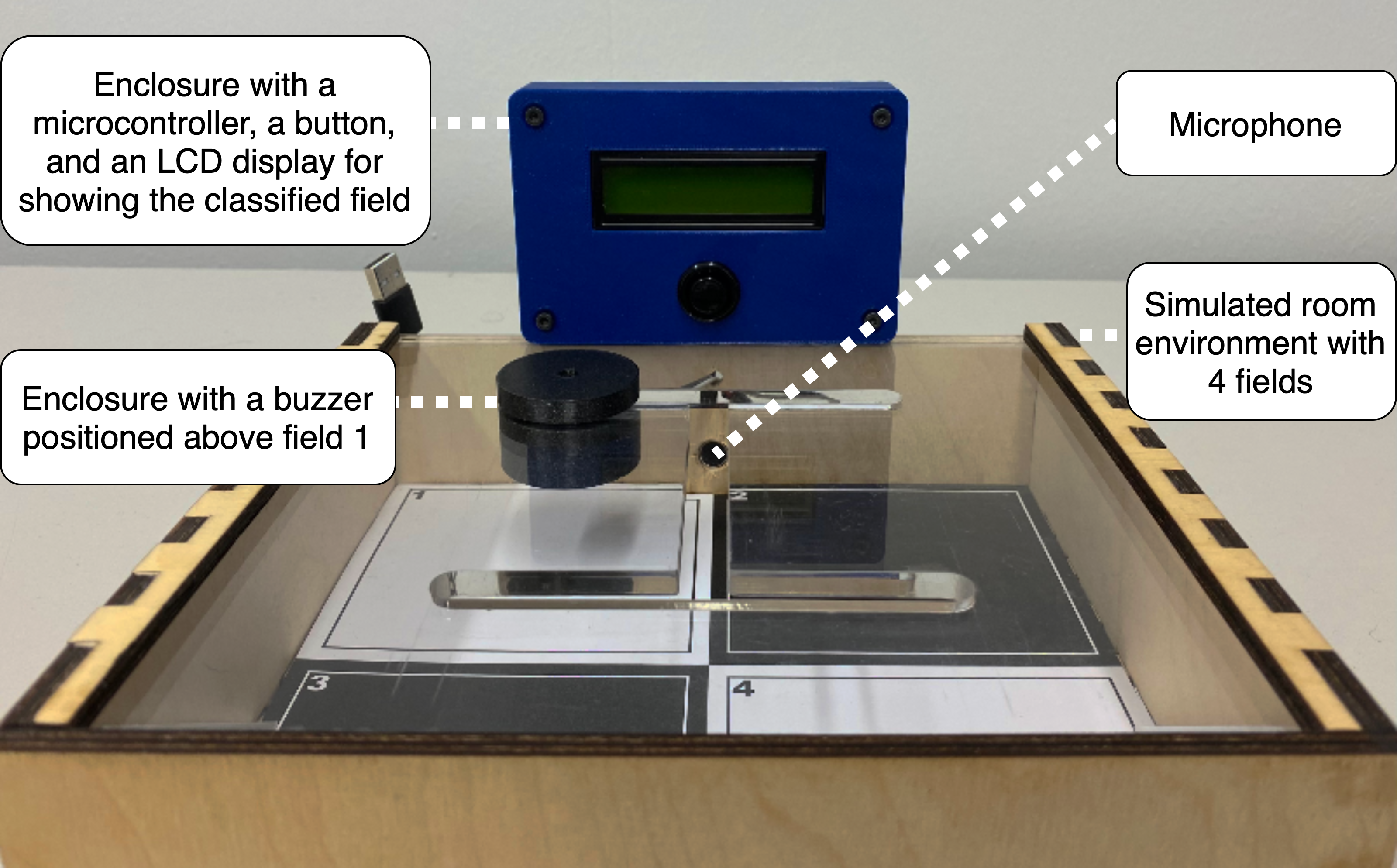}
  \caption{Sound localization demonstrator.}
  \label{fig:SoundLocalizationDemonstrator}
\end{figure}

The system concept presented in this paper is not only applicable to the scenario of determining the fill levels of waste containers, but it also extends to numerous other application areas. For instance, we have successfully applied the entire concept to smaller containers, achieving also a detection accuracy exceeding 90\%, in an attempt to determine their fill levels. The experimental setup for this application is depicted in \Cref{fig:FillLevelDemonstrator}. Furthermore, the concept has been effectively employed in the field of sound localization. In our experimental setup for this application, four marked areas were arranged within a small box designed to simulate a room. Depending on the origin of the sound playback or where the RIR is calculated, SIREC can be trained to predict the correct area. The experimental setup for this particular application is illustrated in \Cref{fig:SoundLocalizationDemonstrator}.

\subsection{Validity Considerations and Challenges}
In the context of the described approach in diverse domain areas, including fill level detection and sound localization, it is essential to consider potential threats to validity. One challenge relates to the material composition of the bins or objects being monitored. Variations in material properties could affect sensor readings, potentially leading to inaccuracies in classification. Consequently, the SIREC model should be trained to account for these variations. Furthermore, although noise reduction is achieved through spectral subtraction, external factors such as ambient noise or interference from nearby sources might still introduce disturbances, impacting the system's performance. Additionally, while the experimental setups are designed to simulate real-world conditions, they may not fully capture all the complexities found in realistic environments, possibly limiting the generalizability of the results. Therefore, careful attention to these factors is crucial to ensure the robustness and reliability of our approach across different scenarios and applications. Future work must also take social and privacy aspects into account. Electronic devices with microphones and the possibility of establishing an internet connection can cause a number of security problems if used incorrectly, some of which we would like to briefly address. A frequently cited problem with insecure IoT scenarios is the emergence of botnets. If our approach were to be used in waste management and thus affect a large number of devices, this possibility of attack must be taken into account. Another attack scenario is unauthorized access to the microphones and the resulting impact on privacy. Attackers can originate from all areas of private and public life: surveillance by governments, use of microphones and data by criminals, e.g. for burglaries, or in the private realm, e.g. for stalking.

\section{Conclusion}
\label{sec:Conclusion}
\noindent In our study, we have demonstrated the creative potential of AI for addressing environmental challenges. We focused on utilizing sound and a specifically designed ML algorithm to accurately determine the fill levels of containers. Our AI-supported system classifies these levels with over 90\% accuracy, highlighting its efficacy in assessing waste container volumes. One of the system's major strengths is its incorporation of cost-effective components, which significantly enhance scalability and practical application. Furthermore, we explored the energy consumption associated with deploying AI models on resource-constrained devices. Understanding and optimizing energy consumption ensures prolonged battery life, reduced operational costs, and improved sustainability, all of which are vital for edge devices operating in diverse environments with limited power sources. Our approach not only offers a novel method for waste management but also suggests the broader applicability of sound-based classification systems in various scenarios. This methodology could be adapted for other container types, or even applied to entirely different applications and settings, such as surveillance systems for theft protection, thereby broadening the potential applications of sound in classification tasks. Looking forward, we hope our findings will inspire further research into the integration of AI technologies within waste management infrastructures. This study opens up new possibilities for how AI can enhance our environmental engagement and resource stewardship, presenting an exciting trajectory for the role of AI in advancing environmental technology.

\section{Acknowledgment}
The authors would like to thank the students, Mr. Marvin Schacht for his support in constructing the demonstrators, and Mr. Marcel Andres for his assistance with the energy measurements.

\bibliographystyle{IEEEtran} 
\bibliography{BibTeX}

\end{document}